\documentclass{emulateapj}
\usepackage{apjfonts}
 
\usepackage{bm}
\usepackage{color}
\shorttitle{Mean-Field Modeling of $\alpha^2$-Dynamo in Rigidly Rotating Convection}
\shortauthors{Masada and Sano}
\begin{document}
\title{Mean-Field Modeling of $\alpha^2$-Dynamo Coupled with \\Direct Numerical Simulations of Rigidly Rotating Convection}
\author{YOUHEI Masada\altaffilmark{1}, AND TAKAYOSHI Sano\altaffilmark{2}} 
\altaffiltext{1}{Department of Computational Science, Kobe University; Kobe 657-8501, Japan: E-mail: ymasada@harbor.kobe-u.ac.jp}
\altaffiltext{2}{Institute of Laser Engineering, Osaka University; Osaka 565-0871, Japan: E-mail: sano@ile.osaka-u.ac.jp}
\begin{abstract}
The mechanism of large-scale dynamos in rigidly rotating stratified convection is explored by direct numerical simulations (DNS) in 
Cartesian geometry. A mean-field dynamo model is also constructed using turbulent velocity profiles consistently extracted from the 
corresponding DNS results. By quantitative comparison between the DNS and our mean-field model, it is demonstrated that the oscillatory 
$\alpha^2$ dynamo wave, excited and sustained in the convection zone, is responsible for large-scale magnetic activities such as cyclic 
polarity reversal and spatiotemporal migration. The results provide strong evidence that a nonuniformity of the $\alpha$-effect, which is a 
natural outcome of rotating stratified convection, can be an important prerequisite for large-scale stellar dynamos, even without the 
$\Omega$-effect.
\end{abstract}
\section{Introduction} 
The solar magnetism is caused by a large-scale dynamo operating in the solar interior. Its ultimate goal is to reproduce observed 
spatiotemporal evolution of the solar magnetic field, such as cyclic polarity reversals and butterfly-shaped migrations, in the framework 
of magnetohydrodynamics (MHD). Although a growing body of evidence is accumulating to reveal large-scale dynamos in numerical 
MHD models of sun-like stars, unsolved questions remain to be answered if full MHD description of the solar dynamo mechanism is 
to be attained \citep{miesch+09,charbonneau10}.

There are two approaches to simulate stellar dynamo evolution. One is global simulations that comprise the entire volume of 
a convection layer and the other uses local-box calculations of a small patch of the stellar interior. The first simulation of a global dynamo 
that succeeded in obtaining solar-like cyclic large-scale magnetic fields was performed by 
\citet{ghizaru+10} \citep[see also][]{brown+11,nelson+13, masada+13}. Another pioneering global simulation was done by \citet{kapyla+12}, 
which reproduced solar-like butterfly-shaped migration of magnetic activity belts. However, a definitive explanation on what regulates the 
solar-like magnetic cycle has yet to be obtained \citep[e.g.,][]{simard+13,kapyla+13b}. The complicated processes included in global 
simulations often preclude elucidating the real essence of the large-scale dynamo.

Direct numerical simulations (DNS) of the convective dynamo in local Cartesian geometry is complementary to the global model, and is 
expected to facilitate our knowledge of the nature of convective dynamos \citep[e.g.,][and references therein]{cattaneo+06,favier+13}. 
Similar to global simulations, the oscillatory large-scale dynamo can also be seen in the Cartesian geometry for rigidly rotating 
stratified convection \citep[][]{kapyla+13a,masada+14a}. Since mean velocity shear is absent in this system, only a stochastic process 
due to turbulent convection would contribute to the large-scale dynamo \citep[e.g.,][]{baryshnikova+87,radler+87}. As a milestone toward 
a complete understanding of the solar MHD dynamo, the mechanism underlying the large-scale dynamo in the Cartesian models must 
be specified. 

In this Letter, we quantitatively demonstrate that the $\alpha^2$ mechanism is responsible for the quasi-periodic features of this 
large-scale dynamo. First, using a DNS model in Cartesian geometry, we investigate the characteristic behaviors of the convective 
dynamo. Next, to disentangle the complicated MHD dynamo processes, we construct a mean-field (MF) $\alpha^2$ dynamo model, using the 
DNS results as profiles of the convective turbulent velocity and helicity. Our MF model is tested by assessing the dependence of the 
dynamo properties on the magnetic diffusivity. Through careful comparison between DNS results and our MF modeling, the mechanism 
underlying large-scale dynamo is revealed.
\section{Large-scale Dynamo in the Reference Model}
We use the same model (model B) as studied by \citet{masada+14a} (hereafter MS14a) as a reference model, in which the large-scale 
dynamo was successfully operated. What follows is a brief review of our numerical MHD model.

In MS14a, a convective dynamo was solved by Cartesian domain (see Figure 1a). This computational domain comprises three layers: 
a top cooling layer (depth $0.15d$,), a middle convection layer (depth $d$), and a bottom stably stratified layer of depth $0.85d$. The 
horizontal size is assumed to be $4d$ (in $x$) $\times\ 4d$ (in $y$). The basic equations are compressible MHD equations in the 
rotating frame of reference, with a constant angular velocity $\bm{\Omega} = -\Omega_0 \bm{e}_z$. 

The initial hydrostatic balance is described by a polytropic distribution with the polytropic index $m$,
\begin{equation}
{\rm d} \epsilon/{\rm d} z = g_0/[(\gamma-1)(m + 1)] \;,
\end{equation}
where $\epsilon$ is the specific internal energy, $\gamma$ is the adiabatic index, and $g_0$ is the constant gravity. Here, when $m < 1.5$, 
it becomes convectively unstable. We choose $m=1$ 
for the convection zone, and $m=3$ for the stable zone. The density contrast between the top and bottom of the domain is $ \simeq 10$. 

The dimensionless quantities are introduced by setting $d=g_0=\rho_0=1$, where $\rho_0$ is the initial density at the top surface. 
The units of length, time, velocity, and magnetic field are then given by $d$, $\sqrt{d/g_0}$, $\sqrt{d g_0}$, and $\sqrt{d g_0\rho_0}$, 
respectively. The volume average in the convection zone and the horizontal average are denoted by single angular brackets containing 
the subscripts ``\rm v" and ``\rm h", respectively. The time-average of each spatial mean is denoted by an additional set of angular brackets. 
The mean convective velocity and the equipartition field strength are defined by 
$u_{\rm cv} \equiv \sqrt{\langle \langle u_z^2 \rangle\rangle_{\rm v}}$ and $B_{\rm cv} \equiv \sqrt{\langle \langle \rho {\bm u}^2 \rangle\rangle_{\rm v}} $. 
The Coriolis number and the convective turnover time are subsequently given by ${\rm Co} = 2\Omega_0d/u_{\rm cv}$ and 
$\tau_{\rm cv} \equiv d/u_{\rm cv}$. 

All the variables are periodic in the horizontal direction, whereas stress-free boundary conditions are used in the vertical direction for the 
velocity. Perfect conductor and vertical field conditions are used for the magnetic field at the bottom and top boundaries, respectively. 
While a constant energy flux is imposed at the bottom boundary, the internal energy remains fixed at the top boundary. 

The fundamental equations are solved by the second-order Godunov-type finite-difference scheme that employs an approximate MHD 
Riemann solver \citep{sano+98}. The magnetic field evolves with the Consistent MoC-CT method \citep{evans+88,clarke96}. 
Non-dimensional parameters of $Pr=1.4$ (Prandtl number), $Pm=4$ (magnetic Prandtl number), and $Ra = 4\times10^6$ 
(Rayleigh number), constant angular velocity of $\Omega_0 = 0.4$, and the spatial resolution of ($N_x, N_y, N_z$) $=$ $(256,256,128)$ 
were adopted in the reference model (see MS14a for definitions of $Pr$, $Pm$, \& $Ra$).  

\begin{figure}[tpb]
\begin{center}
\scalebox{0.6}{{\includegraphics{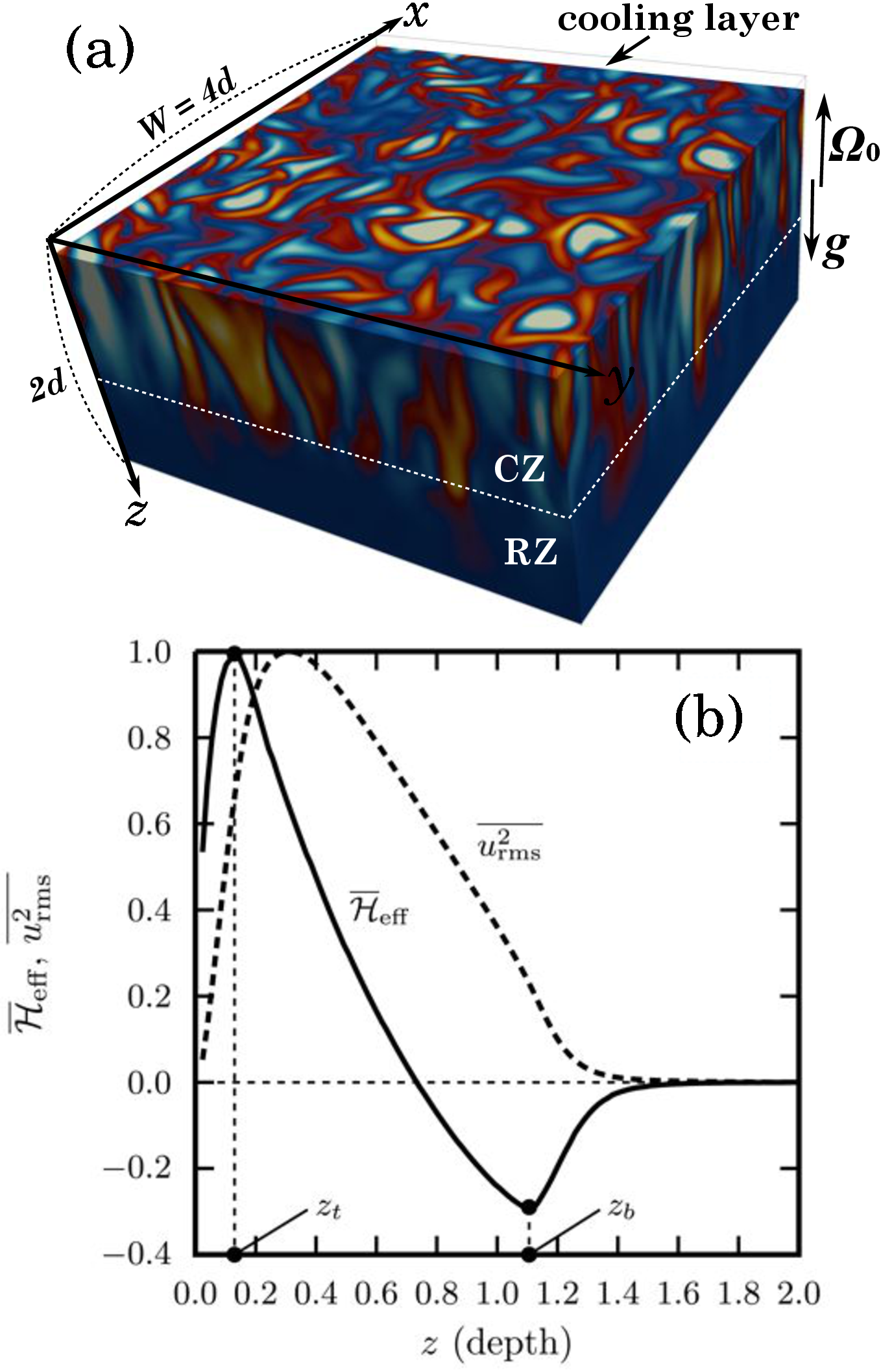}}} 
\end{center}
\caption{(a) Model setup and surface visualization of vertical velocity $u_z$ at $t = 400\tau_{\rm cv}$ for the reference model. 
The red (blue) tone denotes downflow (upflow). (b) The vertical profiles of the effective helicity 
$\mathcal{H}_{\rm eff}$ (solid) and the average velocity $u_{\rm rms}^2$ (dashed). The time average spans in 
the range of $400 \lesssim t/\tau_{\rm cv} \lesssim 800$. The profiles are normalized by their maximum values.}
\label{fig1}
\end{figure}
\begin{figure*}[tbp]
\scalebox{0.6}{{\includegraphics{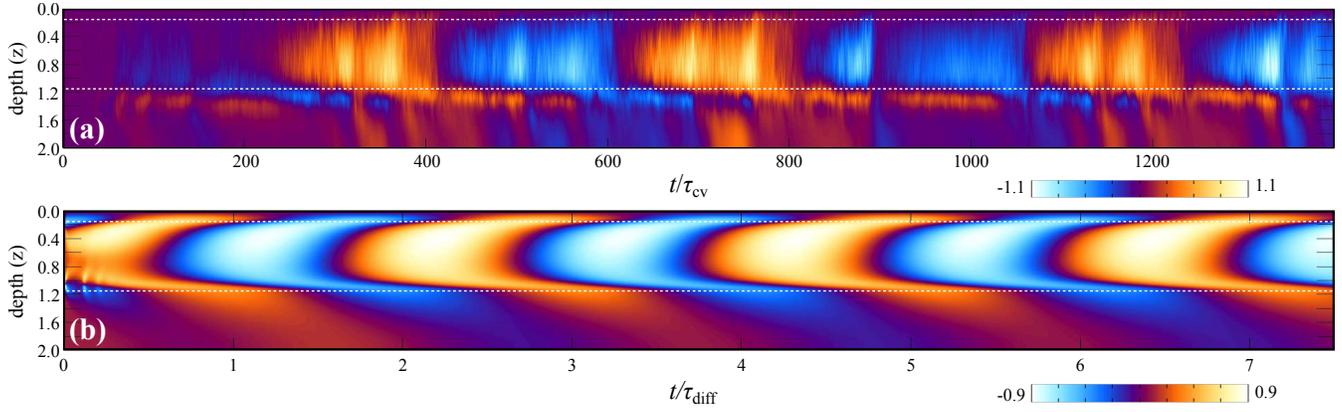}}} 
\caption{Time-depth diagram of $\langle B_x \rangle_{\rm h} $ for the reference model in (a) the DNS and (b) the MF model coupled with 
the DNS. In both panels, the orange (blue) tone denotes the positive (negative) $\langle B_x \rangle_{\rm h} $ in units of $B_{\rm cv}$. 
The horizontal dashed lines show the interface between the convection zone and the stable zones. }
\label{fig2}
\end{figure*}
Initially, small random perturbations are added to the velocity and magnetic fields. Typically after the magnetic diffusion time, a saturated 
turbulent state is achieved. The convective motion there provides $u_{\rm cv} = 0.02$, ${\rm Co} = 40$, $B_{\rm cv} = 0.045$, and 
$\tau_{\rm cv} = 50$. The surface visualization in Figure 1a indicates the vertical velocity at $t = 400\tau_{\rm cv}$, with the red (blue) 
tone denoting downflow (upflow). The convective motion is characterized by cellular upflows surrounded by downflow networks. 
Since there is no symmetry breaking in the horizontal direction, the mean horizontal shear flow, and thus the $\Omega$-effect, are 
absent in our model.

The time-depth diagram of $\langle B_x \rangle_{\rm h}$ is shown in Figure 2a. The orange and blue tones represent positive and negative 
$\langle B_x \rangle_{\rm h} $ in units of $B_{\rm cv}$, respectively. The time is normalized by $\tau_{\rm cv}$. As seen from 
this figure, oscillatory large-scale magnetic field spontaneously organized in our reference model. The $\langle B_x \rangle_{\rm h} $ has a 
peak in the middle of the convection zone and propagates from there to the top and base of the zone. Note that $\langle B_y \rangle_{\rm h}$ 
shows a similar cyclic behavior with $\langle B_x \rangle_{\rm h}$ yet with a phase delay of $\pi/2$ (see also Figure 3). 

It is well known that, in the $\alpha\Omega$ dynamo solution, $B_\phi $ lags $B_r$ by $\pi/4$ (for $\partial \Omega/\partial r > 0$), while 
$B_\phi $ advances $B_r$ by $3\pi/4$ (for $\partial\Omega/\partial r < 0$) \citep[e.g.,][]{brandenburg+05,kapyla+13b}. In contrast, our DNS 
model provides a phase relation similar to the S-parity solution of the MF $\alpha^2$ dynamo model of \citet{brandenburg+09}, wherein 
the vertical field condition is imposed on the top boundary [note that $z$-axis points upward in \citet{brandenburg+09}]. 
If the top perfect conductor condition is adopted in our model, it is expected that $\langle B_y \rangle_{\rm h} $ advances 
$\langle B_x \rangle_{\rm h} $ by $\pi/2$ (A-parity solution).

The large-scale magnetic field with spatiotemporal coherence was a remarkable feature of the convective dynamo achieved in our DNS. 
This feature is reproducible using a mean-field dynamo model with the velocity and helicity profiles consistently extracted from DNS 
results.
\section{Mean-Field Dynamo Model}
\subsection{Governing Equation and Link to DNS}
To explore the mechanism underlying the large-scale dynamo in our DNSs, we construct a one-dimensional MF dynamo model wherein 
velocity profiles are adopted from the DNS results of the saturated convective turbulence and determine the coefficients required for 
MF modeling. See \citet{simard+13} for the similar approach. 

Since the $\Omega$-effect is excluded from our MHD simulations, the $\alpha^2$ dynamo rather than the $\alpha\Omega$ dynamo 
will be realized. The MF equation for the $\alpha^2$ dynamo is obtained from the induction equation, by dividing the variables into the 
horizontal mean values and fluctuating components, $\bm{u} = \langle{\bm{u}}\rangle_{\rm h} + \bm{u}'$ and 
$\bm{B} = \langle \bm{B}\rangle_{\rm h} + \bm{B}'$, and taking the horizontal average:
\begin{equation}
\frac{\partial \langle \bm{B}_h\rangle_{\rm h}}{\partial t} = \nabla \times [ \bm{\mathcal{E}} - \eta_0 \nabla \times  \langle \bm{B}_h\rangle_{\rm h} ] \;,
\end{equation}
with
\begin{equation}
\bm{\mathcal{E}} =  \alpha \langle \bm{B}_h\rangle_{\rm h}  + \gamma {\bm e}_z \times  \langle \bm{B}_h\rangle_{\rm h} - \eta \nabla \times \langle \bm{B}_h\rangle_{\rm h} \;,
\end{equation}
where $\eta_0$ is the microscopic magnetic diffusivity, $\bm{B}_h = (B_x, B_y)$ is the horizontal field, and $\bm{\mathcal{E}}$ is the turbulent 
electromotive force \citep[e.g.,][]{ossendrijver+02}. The coefficients $\alpha$, $\gamma$, and $\eta$ represent the $\alpha$-effect, turbulent 
pumping, and turbulent magnetic diffusivity, respectively. All the terms related to $\langle{\bm{u}}\rangle_{\rm h}$ and $\langle B_z \rangle_{\rm h}$ 
can be ignored in considering the symmetry of the system. All the variables, except for $\eta_0$, depend on the time ($t$) and depth ($z$). 

The MF dynamo described by equation (2) falls into the $\alpha^2$-type category. The MF theory predicts that the $\alpha^2$ mode can 
generate a large-scale magnetic field with an oscillatory nature \citep[e.g.,][]{baryshnikova+87,radler+87,brandenburg+09}. A key ingredient for the oscillatory 
mode is the nonuniformity of the $\alpha$-effect, which can arise naturally as an outcome of rotating stratified convection in the stellar interior. 
Using the rigidly rotating system studied here, the $\alpha^2$ dynamo wave was excited, which propagates only in the depth direction. However, 
as shown by \citet{kapyla+13b}, in the global system, it can travel also in the latitudinal direction due to the strong antisymmetry of the 
$\alpha$-effect across the equator.

The dynamo-generated MF produces a Lorentz force that tends to ``quench" the turbulent motions and control the nonlinear evolution and 
saturation of the system. Since there is no definitive model to describe the magnetic quenching effect \citep[e.g.,][]{rogachevskii+01,blackman+02} 
as yet, we adopt the prototypical models, which are the dynamical $\alpha$-quenching, algebraic $\gamma$- and $\eta$-quenching of the 
catastrophic-type; 
\begin{eqnarray}
\frac{\partial \alpha}{\partial t} &=& - 2\eta_{k} k_c^2 \left[ \frac{\alpha \langle \bm{B}_h\rangle_{\rm h}^2 
- \eta\ (\nabla \times \langle \bm{B}_h \rangle_{\rm h})  \cdot  \langle \bm{B}_h\rangle_{\rm h}}{B_{\rm eq}^2} + \frac{\alpha - \alpha_k}{Re_M}\right] \;, \\
\gamma  & = & \frac{\gamma_k}{ 1 + Re_M \langle \bm{B}_h\rangle_{\rm h}^2/B_{\rm eq}^2}  \;,  \\
\eta  & = & \frac{\eta_{k}}{1 + Re_M \langle \bm{B}_h\rangle_{\rm h}^2/B_{\rm eq}^2} \;, 
\end{eqnarray}
\citep[see][for the quenching]{brandenburg+05}, where $Re_M = \eta_{k}/\eta_0$. The dependence of the MF model on the 
quenching formula should be discussed in detail in a subsequent paper, however, at least the conclusions of this Letter remain 
independent from the choice of the quenching models. The characteristic wavenumber $k_c$ and the equipartition field strength $B_{\rm eq}$ 
are given by $k_c (z) = 2\pi/H_d$ and $B_{\rm eq} (z) = \langle\langle \rho {\bm u_z}^2 \rangle\rangle_{\rm h}$ in our model, where 
$H_d = - {\rm d}z/{\rm d}\ln\langle\langle \rho \rangle\rangle_{\rm h}$ is the density scale height. Here, the subscript ``$k$" refers to the 
unquenched coefficient, which is calculated from DNS results of the saturated convective turbulence.

In the first-order smoothing approximation (FOSA), the unquenched coefficients $\alpha_k$, $\gamma_{k}$ and $\eta_{k}$ in anisotropic 
forms are given by \citep[e.g.,][]{kapyla+06,kapyla+09}, 
\begin{eqnarray}
\alpha_k (z) & = &  - \tau_{c}[ \langle\langle u_z \partial_x u_y \rangle\rangle_{\rm h} 
+ \langle\langle u_x \partial_y u_z\rangle\rangle_{\rm h}] \equiv -\tau_c \mathcal{H}_{\rm eff} \;, \\
\gamma_k (z)  & = & - \tau_{c} \partial_z \langle\langle u_z^2\rangle\rangle_{\rm h} \equiv - \tau_c \partial_z u_{\rm rms}^2\;, \\
\eta_{k} (z)  & = & \tau_{c} \langle\langle u_z^2 \rangle\rangle_{\rm h} \equiv \tau_c u_{\rm rms}^2\;,\ \  
\end{eqnarray}
where $\tau_c$ is the correlation time, $\mathcal{H}_{\rm eff}$ is the effective helicity, and $u_{\rm rms}$ is the root-mean-square velocity. 
The vertical profiles of $\mathcal{H}_{\rm eff}$ and $u_{\rm rms}^2$ in the reference DNS model are shown in Figure 1b by solid and 
dashed lines, respectively. 

The correlation time should be zero in the top cooling and bottom stable layers since the convective turbulence 
is not fully developed; thus $\alpha_k = \gamma_k = \eta_k = 0$ there. Assuming the Strouhal number is unity in the convection zone 
(${\rm St} = \tau_c u_{\rm rms} k_c$ = 1), the vertical profile of $\tau_c$ is given by
\begin{equation}
\tau_{c}(z) = \frac{1}{4u_{\rm rms}k_c} \left[ 1+ {\rm erf}\left( \frac{z-z_b}{h}\right) \right] \left[ 1+ {\rm erf}\left( \frac{z_t-z}{h}\right) \right] \;, 
\end{equation}
where $z_i$ ($i=t, b$) represents the location of the boundaries between regions with and without fully developed turbulence. We define 
$z_t$ and $z_b$ as the depth where $\mathcal{H}_{\rm eff}$ achieves the maximum and minimum values, respectively (see Figure 1b). 
The transition width $h$ is an arbitrary parameter and assumed here as $h = 2\Delta z$ with $\Delta z = 2d/N_z$. The uncertainty of 
$h$ is discussed in the next section. All the coefficients ($\tau_c$, $B_{\rm eq}$, $H_d$, $\alpha_k$, $\gamma_k$, $\eta_k$) required 
for the MF modeling can subsequently be computed from the DNS results.
\subsection{Comparison with DNS}
\begin{figure*}[tb]
\begin{center}
\scalebox{0.6}{{\includegraphics{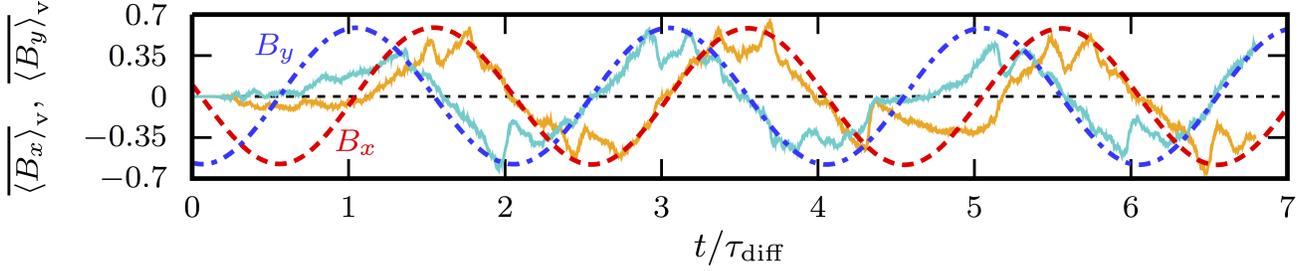}}} 
\caption{The time series of $\langle B_x \rangle_{\rm v}$ and $\langle B_y \rangle_{\rm v}$ for the reference model. The cyan [orange] 
solid line denotes $\langle B_x \rangle_{\rm v}$ [$\langle B_y \rangle_{\rm v}$] normalized by $B_{\rm cv}$ in the DNS. 
The red dashed and blue dash-dotted lines are $\langle B_x \rangle_{\rm v}$ and $\langle B_y \rangle_{\rm v}$ in units of $B_{\rm cv}$ 
in the MF model. The time is normalized by the turbulent magnetic diffusion time.}
\label{fig3}
\end{center}
\end{figure*}
\begin{figure*}[tb]
\begin{center}
\scalebox{0.575}{{\includegraphics{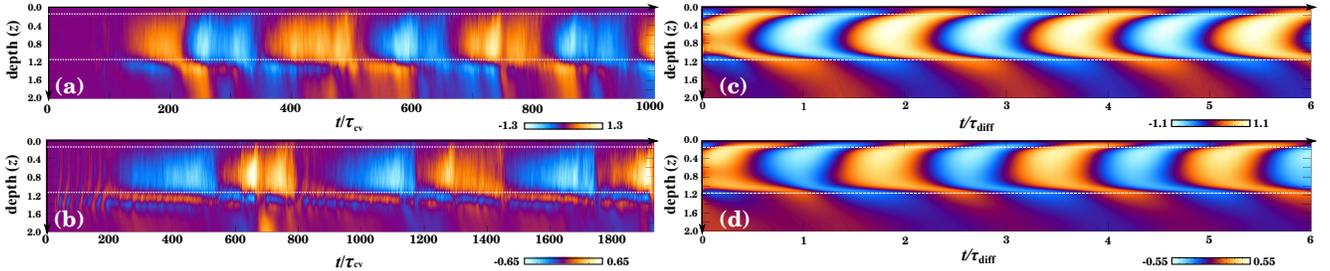}}} 
\caption{Time-depth diagram of $\langle B_x \rangle_{\rm h} $ for the DNSs with (a) $Pm =2$ and (b) $Pm =8$. The MF 
models corresponding to the DNS models with $Pm = 2$ and $8$ are shown in panels (c) and (d). The orange (blue) tone denotes the 
positive (negative) $\langle B_x \rangle_{\rm h} $ in units of $B_{\rm cv}$ in all the panels.}
\label{fig3}
\end{center}
\end{figure*}
Given all the coefficients in equations (2)--(10) from the reference DNS model, the MF equations can be solved using the second-order 
central difference. For time integration, the fourth-order Runge-Kutta method is used. We adopt the same parameters used in the 
DNS: the calculation domain of $0 \le z \le 2d$, the resolution of $N_z = 128$, and the magnetic diffusivity providing $Pm = 4$. 

The time-depth diagram of $\langle B_x \rangle_{\rm h}$ normalized by $B_{\rm cv}$ in the MF model is shown in Figure 2b. The time is 
normalized by turbulent magnetic diffusion time defined by $\tau_{\rm diff} \equiv 1/[\langle\langle \eta \rangle\rangle_{\rm v}k_{d}^2]$, 
where $k_d$ is the typical wavenumber of the dynamo wave and is chosen here as $k_d = \pi/2d$ \citep[c.f.,][]{brandenburg+09}. The 
large-scale field, which is of similar amplitude and spatiotemporal structure as the DNS, is generated and sustained in the bulk of the 
convection zone for the MF model.

Quantitative agreement between the MF model and DNS can be seen in Figure 3, which shows the time series of 
$\langle B_x \rangle_{\rm v}$ and $\langle B_y \rangle_{\rm v}$. The orange [cyan] solid line denotes 
$\langle B_x \rangle_{\rm v}$ [$\langle B_y \rangle_{\rm v}$] normalized by $B_{\rm cv}$ in the DNS and the red dashed [blue dash-dotted] 
line is that in the MF model. The time of the DNS is rescaled by $\tau_{\rm diff}$ with $\langle\langle \eta \rangle\rangle_{\rm v}$ 
evaluated from the MF model and $k_d = \sqrt{\pi}/(2d)$. The longer wavelength required for DNS would be due to the geometrical effect. 
The time of the MF model shifts to match the DNS phase. 

The cycle and amplitude of the large-scale magnetic field in the MF model coincide with those in the DNS. Furthermore, the phase 
difference between $\langle B_x \rangle_{\rm v}$ and $\langle B_y \rangle_{\rm v}$ seen in the DNS model is also reproduced perfectly. 
This indicates that the oscillatory $\alpha^2$ dynamo wave is regulated by the turbulent magnetic diffusivity and is responsible for the 
spatiotemporal evolution of the large-scale magnetic field in the DNS. 
 
\subsection{Validation of our MF Model }
\begin{figure}[tbp]
\begin{center}
\scalebox{0.5}{{\includegraphics{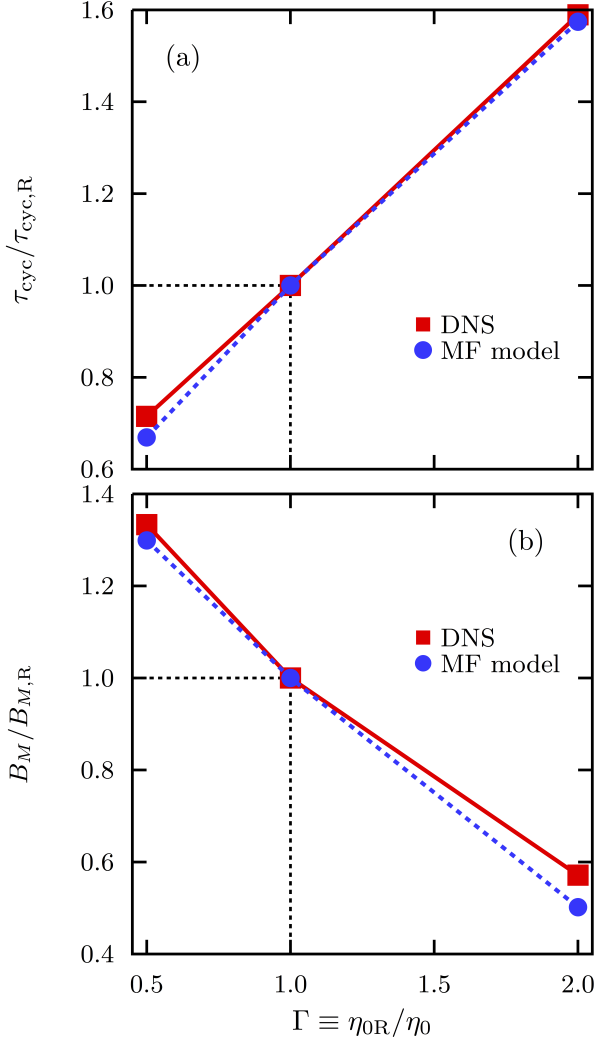}}} 
\caption{
(a) The dynamo period $\tau_{\rm cyc}$ and (b) the saturated field strength 
$B_{\rm M} \equiv [\langle\langle B_x \rangle\rangle_{\rm v}^2 + \langle\langle B_y \rangle\rangle_{\rm v}^2]^{1/2}$ as a 
function of $\Gamma \equiv \eta_{0R}/\eta_0$. 
 The red squares denote the DNS results and the blue circles are the MF models. 
The vertical axis is normalized by the values of the reference model, $\tau_{\rm cyc, R}$ and $B_{\rm M,R}$. } 
\label{fig5}
\end{center}
\end{figure}
To demonstrate the validity of our MF model, we apply it to other DNS models with varying parameters. Here, we focus on the effect of 
magnetic diffusivity ($\eta_0$). The setup is identical to that used in the reference model except for $\eta_0$ or the magnetic Prandtl 
number. The models with $Pm = 2$ and $8$, which adopt two times and half of $\eta_0$ assumed in the reference model, are simulated 
by both DNS and our MF model. Note that $u_{\rm cv} $ and $B_{\rm cv}$ remain unchanged when varying $Pm$. 

The time-depth diagram of $\langle B_x \rangle_{\rm h} $ is shown in Figure 4. Panels (a) and (b) correspond to DNSs with $Pm=2 $ and 
$8$. Regardless of $Pm$, the large-scale oscillatory magnetic field is organized in the bulk of the convection zone. The red squares in 
Figure 5 indicate the $\eta_0$-dependence of (a) the dynamo period $\tau_{\rm cyc}$ and (b) the saturated field strength $B_{\rm M}$, 
where $\tau_{\rm cyc}$ is the statistically averaged value and 
$B_{\rm M} \equiv [\langle\langle B_x \rangle\rangle_{\rm v}^2 + \langle\langle B_y \rangle\rangle_{\rm v}^2]^{1/2}$. 
Each axis is normalized by the value of the reference DNS model ($\eta_{0R},\ \tau_{\rm cyc,R} = 210\tau_{\rm cv},\ B_{\rm M,R} = 0.024$). 
While $\tau_{\rm cyc}$ is inversely proportional to $\eta_0$, $B_{\rm M}$ increases correspondingly. This suggests that magnetic diffusivity 
affects the saturation process of the dynamo in our DNSs.

Following the same procedure as that in \S 3.1, counterpart MF models are constructed. All the coefficients in equations (2)--(10) are 
extracted from the corresponding DNS model. The setup and parameters adopted in the MF models are the same as the reference model, 
except for $\eta_0$. 

Figures 4 (c) and (d) show the time-depth diagram of $\langle B_x \rangle_{\rm h} $ in the MF models corresponding to the DNSs with 
$Pm = 2$ and $8$. Time is normalized by $\tau_{\rm diff}$ with $\langle\langle \eta \rangle\rangle_{\rm v}$ evaluated from each MF 
model. Evidently, the similar spatiotemporal structure of the large-scale field with the DNS is also built up in the MF model. The blue 
circles in Figure 5 represent $\tau_{\rm cyc}$ and $B_{\rm M}$ for the MF model. Normalization units are those of the reference MF 
model, $\tau_{\rm cyc,R} = 3380 \sqrt{d/g_0}$ and $ \ B_{\rm M,R} = 0.028$. The slope and amplitude of the $\eta_0$-dependence is 
well reproduced by the MF model. These results confirm that the oscillatory large-scale magnetic field observed in the DNS is a 
consequence of the turbulent $\alpha^2$-dynamo.
\section{Summary \& Discussion}
In this Letter, the mechanism controlling the large-scale dynamo in rotating stratified convection was examined by DNS in Cartesian 
geometry and the MF dynamo model with the information of turbulent velocity extracted from DNS. We then quantitatively demonstrated 
that the oscillatory $\alpha^2$ dynamo wave, excited and sustained in the convection zone, was responsible for the large-scale dynamo 
with cyclic polarity reversals and spatiotemporal migrations observed in the DNS. Our MF model was validated by evaluating the 
dependence of the large-scale dynamo on the magnetic diffusivity. It is concluded that the nonuniformity of the $\alpha$-effect is a key 
ingredient for the large-scale dynamo with oscillatory nature. 

The oscillatory $\alpha^2$ dynamo mode is attiring a greater level of attention in solar dynamo modeling. Recently, \citet{mitra+10} 
reported an intriguing numerical finding in their forced helical turbulence that $\alpha^2$ dynamo can yield solar-like equatorward 
migration of magnetic activity belts \citep[see also][]{schrinner+11}. The superiority of the $\alpha^2$ mode over $\alpha\Omega$ mode at 
the nonlinear stage was found by \citet{hubbard+11}. Furthermore, a connection between $\alpha^2$ dynamo mode and solar 
magnetism was suggested in some recent results of global MHD dynamos \citep[e.g.,][]{simard+13,kapyla+13b}.  
The crucial factor is the nonuniformity of the $\alpha$-effect. Therefore, accurate numerical modeling of the solar internal $\alpha$ profile 
will offer a way to unveil the mystery of solar magnetism.

There is an arbitrary parameter in the MF model, the thickness of the transition layer $h$ in equation (10). The MF solution is actually 
dependent on this parameter. If the thickness $h$ increases, the cycle period of the MF dynamo becomes shorter and its spatiotemporal 
pattern deviates from that of DNS. The value $h = 2\Delta z$ adopted here is not based on physical reason but is best suited for 
reproducing the spatiotemporal pattern of the large-scale dynamo in the DNS. Several methods, such as imposed- and test-field methods, 
have been previously proposed to directly calculate the MF coefficients from DNS without the use of a statistical turbulence model 
\citep[e.g.,][]{ossendrijver+02,hubbard+09,simard+13}. In contrast, our model is based on FOSA for small-scale turbulence. Not only 
the quenching functions, but also the applicability of FOSA to the anisotropic turbulence remains a matter of debate 
\citep[e.g.,][]{kapyla+06}. Although there is room for improvement in our MF model, it appears to appropriately describe various 
aspects of the large-scale dynamo in the DNS.
\acknowledgments
We are grateful to the anonymous referee for the constructive comments. Computations were carried out on XC30 at NAOJ. This work 
was supported by JSPS KAK-ENHI Grant numbers 24740125 and the joint research project of the Institute of Laser Engineering, Osaka University.

\end{document}